\documentclass[12pt]{article}
\catcode`\@=11
\@addtoreset{equation}{section}

\global\arraycolsep=2pt 
\newcommand{\alg}[1]{\mathfrak{#1}}

\usepackage{mathrsfs,amsbsy,amssymb,latexsym,amsfonts,amsmath,cite}
\usepackage{graphicx,color}
\usepackage{mathtools}

\allowdisplaybreaks

\begin{document}

\begin{flushright}
\parbox{4cm}
{KUNS-2590
\\ \today
} 
\end{flushright}

\vspace*{2cm}

\begin{center}
{\Large \bf 
Yang-Baxter invariance of the Nappi-Witten model
}
\vspace*{1.5cm}\\
{\large 
Hideki Kyono\footnote{E-mail:~h\_kyono@gauge.scphys.kyoto-u.ac.jp} 
and Kentaroh Yoshida\footnote{E-mail:~kyoshida@gauge.scphys.kyoto-u.ac.jp}} 
\end{center}

\vspace*{0.5cm}

\begin{center}
{\it Department of Physics, Kyoto University, \\ 
Kitashirakawa Oiwake-cho, Kyoto 606-8502, Japan} 
\end{center}

\vspace{1cm}

\begin{abstract}
We study Yang-Baxter deformations of the Nappi-Witten model 
with a prescription invented by Delduc, Magro and Vicedo.  
The deformations are specified by skew-symmetric classical $r$-matrices 
satisfying (modified) classical Yang-Baxter equations. 
We show that the sigma-model metric is invariant under arbitrary deformations 
(while the coefficient of $B$-field is changed) by utilizing the most general classical $r$-matrix. 
Furthermore, the coefficient of $B$-field is determined to be the original value from the requirement 
that the one-loop $\beta$-function should vanish. 
After all, the Nappi-Witten model is the unique conformal theory within the class of 
the Yang-Baxter deformations preserving the conformal invariance.   
\end{abstract}

\setcounter{footnote}{0}
\setcounter{page}{0}
\thispagestyle{empty}

\newpage

\tableofcontents

\section{Introduction}

The Yang-Baxter sigma-model description, which was originally proposed by Klimcik \cite{Klimcik}, 
is a systematic way to consider integrable deformations of 2D non-linear sigma models. 
According to this procedure, the deformations are specified by skew-symmetric classical 
$r$-matrices satisfying the modified classical Yang-Baxter equation (mCYBE)\,. 
The original work \cite{Klimcik} has been generalized to symmetric spaces \cite{DMV} 
and the homogeneous CYBE \cite{MY-YBE}. 

\medskip 

Yang-Baxter deformations of the AdS$_5\times$S$^5$ superstring can be studied  
with the mCYBE \cite{DMV2} and the CYBE \cite{KMY-Jordanian-typeIIB}. 
For the former case, the metric and $B$-field are derived in \cite{ABF} 
and the full background has recently been studied in \cite{ABF2,HT}. 
For the latter case, classical $r$-matrices are identified with solutions 
of type IIB supergravity including $\gamma$-deformations of S$^5$ \cite{LM,Frolov}
and gravity duals of non-commutative gauge theories \cite{HI,MR},  
in a series of works \cite{LM-MY,MR-MY,Sch-MY,SUGRA-KMY,MY-duality,Stijn,KKSY,CMY} 
(For a short summary, see \cite{MY-summary}). 

\medskip 

Lately, Yang-Baxter deformations of 4D Minkowski spacetime have been studied \cite{MORSY,BKLSY}. 
In \cite{MORSY}, classical $r$-matrices are identified with exactly-solvable string backgrounds 
such as Melvin backgrounds and pp-wave backgrounds. In \cite{BKLSY}, 
Yang-Baxter deformations of 4D Minkowski spacetime are discussed by using classical $r$-matrices 
associated with $\kappa$-deformations of the Poincar\'e algebra \cite{kappa}. 
Then the resulting deformed geometries include T-duals of (A)dS$_4$ spaces\footnote{T-dual of dS$_4$ 
can be derived as a scaling limit of $\eta$-deformed AdS$_5$ as well \cite{PT}.}   
and a time-dependent pp-wave background. 
Furthermore, the Lax pair is presented for the general $\kappa$-deformations \cite{BKLSY,KSY}. 

\medskip 

As a spin off from this progress, it would be interesting to study Yang-Baxter deformations 
of the Nappi-Witten model \cite{NW}. The target space of this model is given 
by a centrally extended 2D Poincar\'e group. Hence the Yang-Baxter deformed 
Nappi-Witten models can be regarded as toy models of the previous works \cite{MORSY,BKLSY}, 
because the structure of the target space is much simpler than that of 4D Minkowski spacetime. 
This simplification makes it possible to study the most general Yang-Baxter deformation. 
As a matter of course, it is exceedingly complicated in general, hence such an analysis has not been done yet. 

\medskip 

In this article, we investigate Yang-Baxter deformations of the Nappi-Witten model 
by following a prescription invented by Delduc, Magro and Vicedo \cite{DMV-WZW}. 
We show that the sigma-model metric is invariant under the deformations 
(while the coefficient of $B$-field is changed) by utilizing the most general classical $r$-matrix.  
Furthermore, the coefficient of $B$-field is determined to be the original value from the requirement 
that the one-loop $\beta$-function should vanish. 
After all, the Nappi-Witten model is the unique conformal theory within the class of 
the Yang-Baxter deformations preserving the conformal invariance (i.\,e., Yang-Baxter invariance). 


\section{Nappi-Witten model}

In this section, we shall give a concise review of the Nappi-Witten model \cite{NW}. 

\medskip 

The Nappi-Witten model is a Wess-Zumino-Witten (WZW) model whose target space is given by 
a centrally extend 2D Poincar\'e group. 
The associated extended Poincar\'e algebra $\mathfrak{g}$ is composed of two translations $P_i~(i=1,2)$\,, 
a rotation $J$ and the center $T$\,. The commutation relations of the generators are given by 
\begin{eqnarray}
\label{2D-P}
[J,P_i]=\epsilon_{ij}P_j\,,\quad[P_i,P_j]=\epsilon_{ij}T\,, \quad 
[T,J] = [T,P_i] = 0\,, 
\end{eqnarray}
where $\epsilon_{ij}$ is an anti-symmetric tensor normalized as $\epsilon_{12}=1$\,. 
It is convenient to introduce a notation of the generators with the group index $I$ 
like 
\begin{eqnarray}
T_I=\bigl\{P_1,P_2,J,T\bigr\} \qquad (I=1,2,3,4)\,.
\end{eqnarray}

\medskip 

Let us introduce a group element represented by 
\begin{eqnarray}
g=\exp\bigl(a_1\,P_1+a_2\,P_2\bigr)\exp\bigl(u\,J+v\,T\bigr)\,. 
\end{eqnarray}
By using this group element $g$\,, the left-invariant current $A$ can be evaluated as  
\begin{eqnarray}
\label{current}
A_\alpha &\equiv& g^{-1}\partial_\alpha g
= A^I\, T_I
\, \nonumber \\
&=&\left(\cos u\, \partial_\alpha a_1+\sin u\, \partial_\alpha a_2\right)\,P_1
+\left(\cos u\, \partial_\alpha a_2-\sin u\, \partial_\alpha a_1\right)\,P_2\nonumber\\
&&+\partial_\alpha u\, J+\left[\partial_\alpha v 
+ \frac{1}{2}\,a_2\, \partial_\alpha a_1-\frac{1}{2}\,a_1\,\partial_\alpha a_2\right]\,T\,.
\end{eqnarray}
Here the index $\alpha=\tau$\,, $\sigma$ is for the world-sheet coordinates. 
It is also helpful to introduce the light-cone expression of $A$ on the world-sheet like 
\begin{eqnarray}
A_\pm\equiv A_\tau \pm A_\sigma\,. 
\end{eqnarray}
By using $A_{\pm}$\,, the classical action of the Nappi-Witten model is given by
\begin{eqnarray}
\label{NWaction}
S[A]=\frac{1}{2}\int_\Sigma\! d^2\sigma ~ \Omega_{IJ}\, A^I_-A^J_+ 
+ \frac{1}{6}\int_{B_3}\!\! d^3\sigma~\epsilon^{\hat{\alpha}\hat{\beta}\hat{\gamma}}\,
\Omega_{KL}\, {f_{IJ}}^LA^{I}_{\hat{\alpha}}A^{J}_{\hat{\beta}}A^{K}_{\hat{\gamma}}\,.
\end{eqnarray}
This action is basically composed of the two parts, 1) the sigma model part and 
2) the Wess-Zumino-Witten (WZW) term. 

\medskip 

The sigma model part is defined as usual on the world sheet $\Sigma$\,, 
where we assume that $\Sigma$ is compact and the periodic boundary condition 
is imposed for the dynamical variables. 
A key ingredient contained in this part is 
the most general symmetric two-form\footnote{The overall factor of $\Omega_{IJ}$ 
(i.e., the level of the WZW model) 
is set to be 1 because it is irrelevant to the deformations we consider later.}
\begin{eqnarray}
\label{symf}
\Omega_{IJ} \equiv 
\begin{pmatrix}
1&~0~&~0~&0\;\\
0&1&0&0\\
0&0&b&1\\
0&0&1&0 
\end{pmatrix}\,, 
\end{eqnarray}
which satisfies the following condition: 
\begin{eqnarray}
\label{invariance}
{f_{IJ}}^K\Omega_{LK}+{f_{IL}}^K\Omega_{JK}=0\,. 
\end{eqnarray}
Here ${f_{IJ}}^K$ are the structure constants which determine the commutation relations 
\[
[T_I,T_J]={f_{IJ}}^KT_K\,.
\]

\medskip

The WZW term in (\ref{NWaction}) also contains $\Omega_{IJ}$\,, but, 
apart from this point, it is the same as the usual.   
The symbol $B_3$ denotes a 3D space which has $\Sigma$ as a boundary. 
Hence the domain of $A^I$ is implicitly generalized to $B_3$ 
with $\hat{\alpha}=\tau\,,\sigma$ and $\xi$ in the WZW term\footnote{
For the detail of the WZW model, for example, see \cite{AAR}.}, 
where the extra direction is labeled by $\xi$\,.  

\medskip 

A remarkable point is that the action (\ref{NWaction}) can be rewritten 
into the following form \cite{NW}:\footnote{In this derivation, we have used the identities (12) and (13) in \cite{NW}.  }
\begin{eqnarray}
\label{NWaction2}
S=-\frac{1}{2}\int_{\Sigma}\! d^2\sigma\,  \Bigl[\,
\gamma^{\alpha\beta}g_{\mu\nu}\,\partial_\alpha X^\mu \partial_\beta X^\nu 
-\epsilon^{\alpha\beta}\, B_{\mu\nu}\, \partial_{\alpha} X^\mu\partial_{\beta} X^\nu
\Bigr]\,.
\end{eqnarray}
Here $X^\mu=\{u,v,a_1,a_2\}$ are the dynamical variables. 
The metric and anti-symmetric two-form on $\Sigma$ are described by 
$\gamma^{\alpha\beta}=\text{diag}(-1,1)$ and $\epsilon^{\alpha\beta}$ 
normalized as $\epsilon^{\tau\sigma}=1$\,. 
Then the space-time metric $g_{\mu\nu}$ and two-form field $B$ are given by 
\begin{eqnarray}
\label{NW-metric}
ds^2&=&g_{\mu\nu}\,dX^\mu dX^\nu
=2dudv+b\,du^2 +da_1^2+da_2^2+a_1\,da_2du-a_2\,da_1du\,,\nonumber\\
B&=&B_{\mu\nu}\,dX^{\mu}\wedge dX^{\nu} =u~da_1 \wedge da_2\,.
\end{eqnarray}
This is a simple 4D background. 

\medskip 

It would be helpful to further rewrite the background (\ref{NW-metric})\,. 
By performing the following coordinate transformation with a real constant $m$ \cite{Gibbons}
\begin{eqnarray}
a_1~~&\rightarrow&~~ a_1\,\cos (m\,u) + a_2\,\sin( m\,u)\,,\qquad 
a_2~~\rightarrow~~ a_1\,\sin( m\,u) - a_2\, \cos( m\,u)\,,\nonumber\\
u ~~&\rightarrow&~~ 2m\,u\,,\qquad v\rightarrow \frac{1}{2m}\,v-b\,m\,u\,, 
\end{eqnarray}
the background (\ref{NW-metric}) can be rewritten as 
\begin{eqnarray}
\label{NW pp-wave}
ds^2&=& 2dudv-m^2(a_1^2+a_2^2)\,du^2 +  da_1^2 + da_2^2\,,\nonumber\\
B&=&-2m\,u~da_1 \wedge da_2+2m^2a_1\,u~da_1 \wedge du+2m^2a_2\,u~da_2 \wedge du\,.
\end{eqnarray}
This is nothing but a pp-wave background. 
Note here that the last two terms of $B$-field in (\ref{NW pp-wave}) contribute 
to the Lagrangian as the total derivatives, which can be ignored in the present setup. 
For this background (\ref{NW pp-wave}), the world-sheet $\beta$-function vanishes 
at the one-loop level\cite{NW}.

\section{Yang-Baxter deformed Nappi-Witten model}

In this section, let us consider Yang-Baxter deformations of the Nappi-Witten model. 

\subsection{A Yang-Baxter deformed classical action}

As explained in the previous section, the Nappi-Witten model contains the WZW term.  
Hence it is not straightforward to study Yang-Baxter deformations of this model. 
Our strategy here is to follow a prescription invented by Delduc, Magro and Vicedo \cite{DMV-WZW}. 
This is basically a two-parameter deformation. It is an easy task to extend their prescription 
to the Nappi-Witten model. 

\medskip 

A deformed action we propose is the following:\footnote{Here we would like to start 
from the WZNW model (non-conformal) 
rather than the WZW model (conformal).  
Hence, a real constant $k$ has been put 
in front of the WZW term to measure the non-conformality.} 
\begin{eqnarray}
\label{deformed action}
S &=& \frac{1}{2}\int_\Sigma\! d^2\sigma ~ \Omega_{IJ}\,A^I_-J^J_+ 
+ \frac{1}{2}k\int_{B_3}\!\! d^3\sigma~\Omega_{KL}\,{f_{IJ}}^LA^{I}_{\xi}A^{J}_{-}A^{K}_{+}\,. 
\end{eqnarray}
Here the deformed current $J$ is defined as 
\begin{eqnarray}
J_\pm &\equiv& (1+\omega\, \eta^2)\frac{1\pm\tilde{A}R}{1-\eta^2 R^2}A_\pm\,. 
\end{eqnarray}
First of all, the classical action (\ref{deformed action}) includes three constant parameters $\eta$\,, $\tilde{A}$ and $k$\,. 
The deformation is measured by $\eta$ and $\tilde{A}$\,. The last parameter $k$ 
is regarded as the level. When $\eta=\tilde{A}=0$ and $k=1$\,, 
the action (\ref{deformed action}) is reduced to the original Nappi-Witten model. 

\medskip 

A key ingredient contained in $J$ is a linear operator $R$: $\alg{g}\rightarrow\alg{g}$\,. 
In the context of Yang-Baxter deformations, it is supposed that $R$ should be 
skew-symmetric and satisfy the (modified) Yang-Baxter equation \cite{mCYBE}
\begin{eqnarray}
\label{CYBE}
[R(X),R(Y)]-R([R(X),Y]+[X,R(Y)])=\omega\, [X,Y]\qquad (X,\,Y\in \mathfrak{g})\,.
\end{eqnarray}
The constant parameter $\omega$ can be normalized by rescaling $R$\,, 
hence it is enough to consider the following three cases: $\omega=\pm1$ and 0\,.  
In particular, the case with $\omega =0$ is the homogeneous CYBE.

\subsection{The general solution of the (m)CYBE}

In this subsection, we derive the general solution of the (m)CYBE. 

\medskip 

Let us start from the most general expression of a linear $R$-operator:  
\begin{eqnarray}
\label{R-matrix2}
R(X) &=& M^{IJ}\Omega_{JK}\,x^K\,T_I\,, \qquad 
M^{IJ} \equiv  \sum_i \left( a_i^I\,b_i^J - b_i^I\,a_i^J\right)\,. 
\end{eqnarray}
Here $M^{IJ}$ is an anti-symmetric $4\times 4$ matrix 
which is parametrized as  
\begin{eqnarray}
M^{IJ}=
\begin{pmatrix}
0&~m_1~&~m_2~&m_3\;\\
-m_1&0&m_4&m_5\\
-m_2&-m_4&0&m_6\\
-m_3&-m_5&-m_6&0 
\end{pmatrix}\,, 
\qquad m_i \in \mathbb{R}\,. 
\label{ansatz}
\end{eqnarray}

\medskip 
 
By using the expression (\ref{R-matrix2}) and the defining relation 
of $\Omega_{IJ}$ (\ref{invariance})\,,  
the (m) CYBE (\ref{CYBE}) can be rewritten into the following form:  
\begin{eqnarray}
{f_{LM}}^KM^{LI}M^{MJ}+{f_{LM}}^IM^{LJ}M^{MK}+{f_{LM}}^JM^{LK}M^{MI}
-\omega {f_{LM}}^K\Omega^{LI}\Omega^{MJ}=0\,. \label{cybe-f}
\end{eqnarray}
Note that we define $\Omega^{IJ}$ as the inverse matrix of $\Omega_{IJ}$\,. 
Then, by putting the expression (\ref{ansatz}) into (\ref{cybe-f})\,, 
the most general solution can be determined like 
\begin{eqnarray}
M^{IJ}=
\begin{pmatrix}
0&~\sqrt{\omega}~&~0~&m_3\;\\
-\sqrt{\omega}&0&0&m_5\\
0&0&0&m_6\\
-m_3&-m_5&-m_6&0 
\end{pmatrix}\,.
\label{solution}
\end{eqnarray}
Here the condition (\ref{cybe-f}) has led to the following constraints:
\[
 m_1 = \sqrt{\omega}\,, \qquad m_2 = m_4 =0\,.
\]
Then we have also supposed that $\omega \geq 0$ 
in order to preserve the reality of the background\footnote{When we consider $\omega< 0$\,, 
we need to multiply the linear $R$-operator by the imaginary unit $i$\,.}.
After all, $m_3$\,, $m_5$ and $m_6$ have survived as free parameters of the $R$-operator 
as well as $\omega$\,. 
 
\subsection{The general deformed background}

Let us consider a deformation of the Nappi-Witten model 
with the general solution (\ref{solution})\,.
The resulting background is given by
\begin{eqnarray}
ds^2&=&da_1^2+da_2^2+\frac{(1+\omega\,\eta^2)b-(m_3^2+m_5^2)\eta^2}{1-m_6^2\,\eta^2}du^2
+2\frac{1+\omega\,\eta^2}{1-m_6^2\,\eta^2}dudv 
\nonumber \\
&& + \frac{(1+\omega\,\eta^2)a_2+2\eta^2\{(m_3m_6+m_5\sqrt{\omega})\cos u
-(m_5m_6-m_3\sqrt{\omega})\sin u\}}{1-m_6^2\,\eta^2}da_1du\nonumber\\
&& - \frac{(1+\omega\,\eta^2)a_1-2\eta^2\{(m_5m_6-m_3\sqrt{\omega})\cos u
+(m_3m_6+m_5\sqrt{\omega})\sin u\}}{1-m_6^2\,\eta^2}da_2du\,,\nonumber\\
B &=& k\,u\,da_1\wedge da_2+\frac{\tilde{A}m_6(1+\omega\,\eta^2)}{2(1-m_6^2\eta^2)}
\left(a_2\,da_1\wedge du-a_1\,da_2\wedge du\right)\,. \label{general-bg}
\end{eqnarray}
Here we have ignored the total derivative terms that appeared in the $B$-field part.
This background (\ref{general-bg}) can be simplified 
by performing a coordinate transformation
\begin{eqnarray}
a_1~~&\rightarrow&~~ a_1\,\cos (m\,u) + a_2\,\sin (m\,u) + C_1\cos \left( C_3\,u\right) 
+ C_2\,\sin \left( C_3\,u\right)\,, \nonumber \\
a_2 ~~&\rightarrow&~~ -a_2\,\cos ( m\,u) + a_1\,\sin ( m\,u) 
- C_2\,\cos\left(C_3\,u\right) + C_1\,\sin \left(C_3\,u\right)\,, \nonumber \\
u ~~&\rightarrow&~~ C_3\,u\,, \nonumber \\
v ~~&\rightarrow&~~ \frac{1}{2\,m}v-\frac{1}{2}b\,C_3\,u-\frac{1}{2}\Bigl[C_2\cos \left(\frac{C_3-C_4}{2}u\right) 
-C_1\sin\left(\frac{C_3-C_4}{2}u\right)\Bigr]a_1\nonumber\\
&&+\frac{1}{2}\Bigl[C_1\cos\left(\frac{C_3-C_4}{2}u\right) + C_2\sin\left(\frac{C_3-C_4}{2}u\right)\Bigr]a_2\,, 
\label{trans}
\end{eqnarray}
where we have introduced the following quantities: 
\begin{eqnarray}
C_1 & \equiv &\frac{m_5m_6-m_3\sqrt{\omega}}{m_6^2+\omega}\,,\qquad
C_2 \equiv \frac{m_3m_6+m_5\sqrt{\omega}}{m_6^2+\omega}\,,\nonumber\\
C_3 & \equiv & 2m\frac{1-m_6^2\,\eta^2}{1+\omega\,\eta^2}\,,\qquad
C_4 \equiv 2m\frac{m_6^2+\omega}{1+\omega\,\eta^2}\eta^2\,.
\end{eqnarray}
After performing the transformation (\ref{trans})\,, the resulting background is given by 
the following pp-wave background equipped with a $B$-field: 
\begin{eqnarray}
\label{DBG}
ds^2&=& 2dudv -m^2(a_1^2+a_2^2)du^2 + da_1^2+da_2^2\,,\nonumber\\
B&=&-k\,C_3\,u\,da_1\wedge da_2-m\tilde{A}m_6a_2\,da_1\wedge du+m\tilde{A}m_6 a_1\,da_2\wedge du\,.
\end{eqnarray}
Here we have ignored the total derivative terms again. 

\medskip 

Note that the $B$-field in (\ref{DBG}) can be rewritten as (up to total derivative terms)
\begin{eqnarray}
\label{DBG2}
B&=&-(k\,C_3-2m\,\tilde{A}\,m_6 )\,u\,da_1\wedge da_2\,.
\end{eqnarray}
Comparing (\ref{DBG2}) with the $B$-field in (\ref{NW pp-wave})\,, 
one can find that only the difference is the coefficient of $B$-field. 
From the viewpoint of the original Nappi-Witten model, the coefficient of the WZW term 
has been changed and the resulting theory should be regarded as a WZNW model.   
According to this observation, it is obvious that the deformed model is exactly solvable\footnote{
In the case of \cite{DMV-WZW}, it is necessary to impose a condition for $\tilde{A}$ 
so as to preserve the integrability (c.f., \cite{KOY})\,. 
However, such an extra condition is not needed in the present case. }.

\subsection{Yang-Baxter deformations and conformal invariance}

Finally, let us show that the original Nappi-Witten model is the unique conformal theory 
within the class of the Yang-Baxter deformations preserving the conformal invariance. 


\medskip 

Due to the requirement of the vanishing $\beta$-function at the one-loop level, 
the two $B$-fields should be identical as follows: 
\begin{eqnarray}
2m \quad (\mbox{the original}) ~~=~~  k\,C_3-2m\,\tilde{A}\,m_6 \quad (\mbox{the deformed})\,.  
\label{equal}
\end{eqnarray} 
%
%
This condition indicates two interesting results. The first one is the Yang-Baxter invariance of 
the Nappi-Witten model. 
If we start from the case with $k=1$\,, 
then the original system is invariant under the Yang-Baxter deformations preserving the conformal 
invariance, which are specified by the parameters satisfying the condition 
\[
1 = \frac{1+\omega\,\eta^2}{1-m_6^2\,\eta^2}(1+m_6\tilde{A})\,.
\]
In other words, the Yang-Baxter invariance follows from the conformal invariance. 

\medskip 

The second is that the Yang-Baxter deformation may map a non-conformal theory to 
the conformal Nappi-Witten model. Suppose that we start from the case with $k \neq 1$\,. 
Then, by performing a Yang-Baxter deformation with parameters satisfying the condition 
\begin{eqnarray}
k=\frac{1+\omega\,\eta^2}{1-m_6^2\,\eta^2}(1+m_6\tilde{A})\,, 
\end{eqnarray}
the resulting system becomes the Nappi-Witten model. 
In other words, the coefficient of $B$-field can be set to the conformal fixed point 
by an appropriate Yang-Baxter deformation.

\section{Conclusion and discussion}

In this article, we have studied Yang-Baxter deformations of the Nappi-Witten model. 
By considering the most general classical $r$-matrix, 
we have shown the invariance of the sigma-model metric under arbitrary deformations, 
up to two-form $B$-fields. That is, the effect coming from the deformations 
is reflected only as the coefficient of $B$-field. Then, the coefficient of $B$-field has been determined 
to be the original value from the requirement that the one-loop $\beta$-function should vanish. 
After all, it has been shown that the Nappi-Witten model is the unique conformal theory within the class of 
the Yang-Baxter deformations preserving the conformal invariance (i.\,e., Yang-Baxter invariance).


\medskip 

There are many future directions. It would be interesting to consider a supersymmetric 
extension of our analysis by following \cite{super-NW}. The number of the remaining supersymmetries 
should depend on Yang-Baxter deformations because the coefficient of $B$-field is changed. 
It is also nice to investigate higher-dimensional cases (e.g., 
the maximally supersymmetric pp-wave background \cite{BFHP}). As another direction, 
one may consider non-relativistic backgrounds such as Schr\"odinger spacetimes \cite{Sch}
and Lifshitz spacetimes \cite{Lifshitz}. Although there is a problem 
of the degenerate Killing form similarly, it can be resolved by adopting the most general 
symmetric two-form \cite{SYY} as in the Nappi-Witten model. 
It would be straightforward to apply the techniques 
presented in \cite{SYY} to Yang-Baxter deformations by following our present analysis.  

\medskip 

It should be remarked that the most interesting indication of this work is 
the universal aspect of the dual gauge-theory side. According to our work, 
pp-wave backgrounds would have a kind of rigidity against Yang-Baxter deformations. 
This result may indicate that the ground state and lower-lying excited states of 
the spin chain associated with the $\mathcal{N}=4$ super Yang-Mills theory are {\it invariant}. 
It is quite significant to extract such a universal characteristic after classifying various examples. 
This is the standard strategy in theoretical physics and would be much more important than 
identifying the associated dual gauge theory for each of the deformations.

\medskip 

We hope that our result could shed light on a universal aspect of Yang-Baxter deformations 
from the viewpoint of the invariant pp-wave geometry.

\subsection*{Acknowledgments}

We are very grateful to Martin Heinze and Jun-ichi Sakamoto for useful discussions. 
The work of K.Y. is supported by Supporting Program for Interaction-based Initiative Team Studies 
(SPIRITS) from Kyoto University and by the JSPS Grant-in-Aid for Scientific Research (C) No.15K05051.
This work is also supported in part by the JSPS Japan-Russia Research Cooperative Program 
and the JSPS Japan-Hungary Research Cooperative Program.


\begin{thebibliography}{99}

\bibitem{Klimcik}
 C.~Klimcik,
  ``Yang-Baxter sigma models and dS/AdS T duality,''  
JHEP {\bf 0212} (2002) 051  [hep-th/0210095]; 
  ``On integrability of the Yang-Baxter sigma-model,''  
J.\ Math.\ Phys.\  {\bf 50} (2009) 043508  [arXiv:0802.3518 [hep-th]]; 
 ``Integrability of the bi-Yang-Baxter sigma model,''  Lett.\ Math.\ Phys.\  {\bf 104} (2014) 1095
  [arXiv:1402.2105 [math-ph]].  

\bibitem{DMV}
  F.~Delduc, M.~Magro and B.~Vicedo,
  ``On classical q-deformations of integrable sigma-models,''  
  JHEP {\bf 1311} (2013) 192  [arXiv:1308.3581 [hep-th]].   

\bibitem{MY-YBE}
 T.~Matsumoto and K.~Yoshida,
  ``Yang-Baxter sigma models based on the CYBE,''
  Nucl.\ Phys.\ B {\bf 893} (2015) 287
  [arXiv:1501.03665 [hep-th]].   
  
\bibitem{DMV2}
  F.~Delduc, M.~Magro and B.~Vicedo,
  ``An integrable deformation of the AdS$_5\times$S$^5$ superstring action,''  
 Phys.\ Rev.\ Lett.\  {\bf 112} (2014) 051601
  [arXiv:1309.5850 [hep-th]];   
  ``Derivation of the action and symmetries of the $q$-deformed AdS$_5\times$S$^5$ superstring,'' 
  JHEP {\bf 1410} (2014) 132
  [arXiv:1406.6286 [hep-th]].  
  
\bibitem{KMY-Jordanian-typeIIB}
  I.~Kawaguchi, T.~Matsumoto and K.~Yoshida,
  ``Jordanian deformations of the AdS$_5\times$S$^5$ superstring,''
  JHEP {\bf 1404} (2014) 153
  [arXiv:1401.4855 [hep-th]].
      
\bibitem{ABF}
    G.~Arutyunov, R.~Borsato and S.~Frolov,
  ``S-matrix for strings on $\eta$-deformed AdS$_5\times$S$^5$,'' 
  JHEP {\bf 1404} (2014) 002 [arXiv:1312.3542 [hep-th]].

\bibitem{ABF2}
G.~Arutyunov, R.~Borsato and S.~Frolov,
  ``Puzzles of $\eta$-deformed AdS$_5\times$S$^5$\,,'' 
JHEP {\bf 1512} (2015) 049 [arXiv:1507.04239 [hep-th]]. 


\bibitem{HT}
 B.~Hoare and A.~A.~Tseytlin,
  ``Type IIB supergravity solution for the T-dual of the $\eta$-deformed AdS$_{5} \times$ S$^{5}$ superstring,''
  JHEP {\bf 1510} (2015) 060
  [arXiv:1508.01150 [hep-th]].   
  
\bibitem{LM}
 O.~Lunin and J.~M.~Maldacena,
  ``Deforming field theories with $U(1) \times U(1)$ global symmetry and their gravity duals,''  
JHEP {\bf 0505} (2005) 033  [hep-th/0502086].

\bibitem{Frolov}
   S.~Frolov,
   ``Lax pair for strings in Lunin-Maldacena background,''
   JHEP {\bf 0505} (2005) 069
   [hep-th/0503201].
  
\bibitem{HI}
  A.~Hashimoto and N.~Itzhaki,
  ``Noncommutative Yang-Mills and the AdS / CFT correspondence,''
  Phys.\ Lett.\ B {\bf 465} (1999) 142
  [hep-th/9907166].

\bibitem{MR}
  J.~M.~Maldacena and J.~G.~Russo,
  ``Large N limit of noncommutative gauge theories,''
  JHEP {\bf 9909} (1999) 025
  [hep-th/9908134].    
  
\bibitem{LM-MY}
  T.~Matsumoto and K.~Yoshida,
  ``Lunin-Maldacena backgrounds from the classical Yang-Baxter equation 
- towards the gravity/CYBE correspondence,''
  JHEP {\bf 1406} (2014) 135
  [arXiv:1404.1838 [hep-th]].    

\bibitem{MR-MY}  
 T.~Matsumoto and K.~Yoshida,
  ``Integrability of classical strings dual for noncommutative gauge theories,''
  JHEP {\bf 1406} (2014) 163 
  [arXiv:1404.3657 [hep-th]].  
  
\bibitem{Sch-MY}
  T.~Matsumoto and K.~Yoshida,
  ``Schr\"odinger geometries arising from Yang-Baxter deformations,'' 
  JHEP {\bf 1504} (2015) 180 [arXiv:1502.00740 [hep-th]].  

\bibitem{SUGRA-KMY}
 I.~Kawaguchi, T.~Matsumoto and K.~Yoshida,
  ``A Jordanian deformation of AdS space in type IIB supergravity,'' 
  JHEP {\bf 1406} (2014) 146 [arXiv:1402.6147 [hep-th]].    

\bibitem{MY-duality}
  T.~Matsumoto and K.~Yoshida,
  ``Yang-Baxter deformations and string dualities,''
  JHEP {\bf 1503} (2015) 137 [arXiv:1412.3658 [hep-th]]. 

\bibitem{Stijn}
S.~J.~van Tongeren,
  ``On classical Yang-Baxter based deformations of the AdS$_5 \times$S$^5$ superstring,''
  JHEP {\bf 1506} (2015) 048 [arXiv:1504.05516 [hep-th]]; 
  ``Yang-Baxter deformations, AdS/CFT, and twist-noncommutative gauge theory,'' 
  Nucl.\ Phys.\ B {\bf 904} (2016) 148 [arXiv:1506.01023 [hep-th]].

  
\bibitem{KKSY} 
  T.~Kameyama, H.~Kyono, J.~Sakamoto and K.~Yoshida,
  ``Lax pairs on Yang-Baxter deformed backgrounds,'' 
  JHEP {\bf 1511} (2015) 043 [arXiv:1509.00173 [hep-th]].  
  
\bibitem{CMY}
   P.~M.~Crichigno, T.~Matsumoto and K.~Yoshida,
   ``Deformations of $T^{1,1}$ as Yang-Baxter sigma models,''
    JHEP {\bf 1412} (2014) 085
   [arXiv:1406.2249 [hep-th]]; 
    ``Towards the gravity/CYBE correspondence beyond integrability 
 -- Yang-Baxter deformations of $T^{1,1}$,'' 
 J.\ Phys.\ Conf.\ Ser.\  {\bf 670} (2016) 1,  012019 [arXiv:1510.00835 [hep-th]].
 
      
\bibitem{MY-summary}
  T.~Matsumoto and K.~Yoshida,
  ``Integrable deformations of the AdS$_{5} \times$S$^5$ superstring 
and the classical Yang-Baxter equation 
{\it -- Towards the gravity/CYBE correspondence --},''
  J.\ Phys.\ Conf.\ Ser.\  {\bf 563} (2014) 1,  012020
  [arXiv:1410.0575 [hep-th]]; 
  ``Towards the gravity/CYBE correspondence -- the current status --,''
  J.\ Phys.\ Conf.\ Ser.\  {\bf 670} (2016) 1,  012033.  
   
\bibitem{MORSY}  T.~Matsumoto, D.~Orlando, S.~Reffert, J.~Sakamoto and K.~Yoshida,
  ``Yang-Baxter deformations of Minkowski spacetime,'' 
  JHEP {\bf 1510} (2015) 185 [arXiv:1505.04553 [hep-th]]. 
 
\bibitem{BKLSY}
A.~Borowiec, H.~Kyono, J.~Lukierski, J.~Sakamoto and K.~Yoshida,
  ``Yang-Baxter sigma models and Lax pairs arising from $\kappa$-Poincar\'e $r$-matrices,''
  arXiv:1510.03083 [hep-th]. 
  
 \bibitem{kappa}
J.~Lukierski, H.~Ruegg, A.~Nowicki and V.~N.~Tolstoy,
  ``Q deformation of Poincare algebra,''
  Phys.\ Lett.\ B {\bf 264} (1991) 331; 
J.~Lukierski and A.~Nowicki,
  ``Quantum deformations of D = 4 Poincare and Weyl algebra from Q deformed D = 4 conformal algebra,''
  Phys.\ Lett.\ B {\bf 279} (1992) 299.

\bibitem{PT}
 A.~Pachol and S.~J.~van Tongeren,
  ``Quantum deformations of the flat space superstring,'' 
  Phys.\ Rev.\ D {\bf 93} (2016) 2,  026008 [arXiv:1510.02389 [hep-th]].

\bibitem{KSY}  
 H.~Kyono, J.~Sakamoto and K.~Yoshida,
  ``Lax pairs for deformed Minkowski spacetimes,'' 
  JHEP {\bf 1601} (2016) 143
  [arXiv:1512.00208 [hep-th]]. 

\bibitem{NW}
C.~R.~Nappi and E.~Witten,
  ``A WZW model based on a nonsemisimple group,''
  Phys.\ Rev.\ Lett.\  {\bf 71} (1993) 3751
  [hep-th/9310112].    

\bibitem{DMV-WZW} 
   F.~Delduc, M.~Magro and B.~Vicedo,
    ``Integrable double deformation of the principal chiral model,''
    Nucl.\ Phys.\ B {\bf 891} (2015) 312
    [arXiv:1410.8066 [hep-th]].    

\bibitem{AAR}
E.~Abdalla, M.~C.~Abdalla and K.~Rothe, 
 ``Non-perturbative methods in two-dimensional quantum field theory,'' 
 World Scientific, 1991. 

\bibitem{Gibbons}
  G.~W.~Gibbons and C.~N.~Pope,
  ``Kohn's Theorem, Larmor's Equivalence Principle and the Newton-Hooke Group,''
  Annals Phys.\  {\bf 326} (2011) 1760
  [arXiv:1010.2455 [hep-th]].

\bibitem{mCYBE}
M.~A.~Semenov-Tyan-Shanskii, ``What is a classical $r$-matrix?'' 
Funct.\ Anal.\ Appl.\ {\bf 17} (1983) 259. 
For a nice review, see ``Integrable Systems and Factorization Problems,'' 
nlin/0209057.   

\bibitem{KOY}
   I.~Kawaguchi, D.~Orlando and K.~Yoshida,
   ``Yangian symmetry in deformed WZNW models on squashed spheres,''
   Phys.\ Lett.\  B {\bf 701} (2011) 475. 
   [arXiv:1104.0738 [hep-th]]; 
   I.~Kawaguchi and K.~Yoshida,
   ``A deformation of quantum affine algebra in squashed WZNW models,'' 
   J.\ Math.\ Phys.\  {\bf 55} (2014) 062302
   [arXiv:1311.4696 [hep-th]].     

\bibitem{super-NW}
  E.~Kiritsis, C.~Kounnas and D.~Lust,
  ``Superstring gravitational wave backgrounds with space-time supersymmetry,''
  Phys.\ Lett.\ B {\bf 331} (1994) 321
  [hep-th/9404114].
  
\bibitem{BFHP}
  M.~Blau, J.~M.~Figueroa-O'Farrill, C.~Hull and G.~Papadopoulos,
  ``A New maximally supersymmetric background of IIB superstring theory,''
  JHEP {\bf 0201} (2002) 047
  [hep-th/0110242].  
  
\bibitem{Sch}
 D.~T.~Son,
  ``Toward an AdS/cold atoms correspondence: A Geometric realization of the Schr\"odinger symmetry,''
  Phys.\ Rev.\ D {\bf 78} (2008) 046003
  [arXiv:0804.3972 [hep-th]];  
 K.~Balasubramanian and J.~McGreevy,
  ``Gravity duals for non-relativistic CFTs,''
  Phys.\ Rev.\ Lett.\  {\bf 101} (2008) 061601
  [arXiv:0804.4053 [hep-th]].
  
\bibitem{Lifshitz}  
S.~Kachru, X.~Liu and M.~Mulligan,
  ``Gravity duals of Lifshitz-like fixed points,''
  Phys.\ Rev.\ D {\bf 78} (2008) 106005
  [arXiv:0808.1725 [hep-th]].

\bibitem{SYY}
  S.~Schafer-Nameki, M.~Yamazaki and K.~Yoshida,
  ``Coset Construction for Duals of Non-relativistic CFTs,''
  JHEP {\bf 0905} (2009) 038
  [arXiv:0903.4245 [hep-th]].    
  
\end{thebibliography}
\end{document}